\documentclass{ws-ijgmmp}
\usepackage{orcidlink}
\usepackage{hyperref}
\begin{document}

\markboth{L.V. Jaybhaye et al.,}
{Constraints on Energy Conditions in $f(R,L_m)$ Gravity}

%
\catchline{}{}{}{}{}
%

\title{Constraints on Energy Conditions in $f(R,L_m)$ Gravity}

\author{Lakhan V. Jaybhaye\orcidlink{0000-0003-1497-276X}}

\address{Department of Mathematics, Birla Institute of Technology and
Science-Pilani,\\ Hyderabad Campus, Hyderabad-500078, India.\\
\email{lakhanjaybhaye@gmail.com} }

\author{Sanjay Mandal\orcidlink{0000-0003-2570-2335}}

\address{Department of Mathematics, Birla Institute of Technology and
Science-Pilani,\\ Hyderabad Campus, Hyderabad-500078, India.\\
\email{sanjaymandal960@gmail.com} }

\author{P. K. Sahoo\orcidlink{0000-0003-2130-8832}}

\address{Department of Mathematics, Birla Institute of Technology and
Science-Pilani,\\ Hyderabad Campus, Hyderabad-500078, India.\\
\email{pksahoo@hyderabad.bits-pilani.ac.in} }

\maketitle

\begin{history}
\received{(19 Nov 2021)}
\revised{(06 Dec 2021)}
\end{history}

\begin{abstract}
In this manuscript, we consider the extension of the Hilbert-Einstein action to analyze several interesting features of the theory. More specifically, the Lagrangian $f(R)$ is replaced by $f(R, L_m)$ in action, where $R$ is the Ricci scalar, and $L_m$ is the matter Lagrangian. We derive the motion equations for a test particle in the Friedmann-Lema\^itre-Robertson-Walker (FLRW) flat and homogeneous spacetime. We also derive the energy conditions in this framework. Then, we use the cosmographic parameter such as Hubble, deceleration, jerk, and snap parameters to constraint the model parameters. As a result, we observe that with the constraint range of model parameters our model shows the current accelerated expansion of the universe.
\end{abstract}



\section{Introduction}\label{sec1}

In the early of 20th century, Albert Einstein proposed general Theory of Relativity, Which changes the perception of the universe. After that, this theory plays a lead role to describe the geodesic nature of spacetime. Later on, it is successfully tested the Solar System tests, but this theory somehow fails to describe all gravity phenomena till date such as accelerated expansion, flatness issue, fine-tuning problem etc. In last few decades, the modern cosmological observations of type Ia supernovae pointed that the expansion of the Universe is accelerating \cite{Riess/1998,Perlmutter/1999,Teqmark/2004,Teqmark/2004a}. According to a modern study, expansion is instancing and examines repercussion of a mysterious force known as Dark Energy, which possesses a crucial negative pressure and positive energy density \cite{Wetterich/1988,Amendola/2000,calwell/2002}. Even so, no incontrovertible evidence has yet materialized to solidify the pressure of Dark Energy. As an outcome, many alternative models have come up to explain this problem.

Moreover, modified theories of gravity are evolved as one of the best way to describe the accelerated expansion of the universe without presence of Dark Energy\cite{Brans,Fara,Dvali,Jacobson,Maattens,Beken}. Some of them are widely known as $f(R)$ gravity theory, $f(T)$ gravity theory, $f(R,T)$ gravity theory, where $R$ is Ricci scalar and $T$ is an energy-momentum tensor \cite{Njiroi/2007,Sotiriou/2010,Capo/2002,Nojiri/2008,Nojiri/2003,Nojiri/2005,Berto/2007,Harko/2008}. The $f(R, L_m)$ gravity theory was proposed recently by including the explicit coupling of the Lagrangian density corresponding to matter ($L_m$) and Ricci scalar ($R$)  \cite{Harko/2010}. Apart from these observations, any modified theory of gravity should follow a significant role in the energy conditions. These conditions define the casual and geodesic composition of spacetime in modified gravity \cite{Kung,Carroll,Allemandi,Berg/2006,Santos/2007,Berto/2009,Wang/2010,Gracia/2011}.

The basic energy conditions used in the general theory of relativity are the weak energy condition (WEC), the null energy condition (NEC), the dominant energy condition (DEC), and the strong energy condition (SEC) \cite{Carroll}. These energy conditions play an essential role in understanding singularity theorem such as black hole thermodynamics \cite{wald}. The Raychaudhuri equations play a vital role in reporting the attractive behavior of gravity and positive energy density \cite{Rau/1955}. There are several approaches in the literature deriving energy conditions by modifying Einstein's GR, one can check for instance, EC constraints in $f(R)$ theory \cite{Santos/2007,Berto/2009}, $f(R,T)$ theory \cite{Moraes/2019}, $f(G)$ theory \cite{Gracia/2011,Bamba/2017}, $f(R,T,R_{\mu\nu}T^{\mu\nu})$ theory \cite{Sharif/2013}, $f(T)$ theory \cite{Liu/2012},$f(R,\mathcal{G})$ theory \cite{Atazadeh/2014}, $f(\mathcal{G},T)$ theory \cite{Sharif/2016},  $f(R,\square R,T)$ gravity \cite{Yousaf/2018}, $f(R,L_m)$ gravity \cite{rl1,rl2},  $f(Q)$ gravity \cite{Mandal}, $f(Q,T)$ gravity \cite{Simran} etc. Also, energy conditions plays a vital role to discuss the casual and geodesic nature of spacetime. Sharif and Zubair did an interesting study in $f(R,L_m)$ gravity, there they discussed nicely second law of thermodynamics in $f(R,L_m)$ gravity \cite{Sharif/2013}. In this paper, we derived and discussed the energy conditions of $f(R,Lm)$ gravity.

The manuscript is presented as in the following steps. In Section-\ref{sec2}, we present the fundamental framework of $f(R, L_m)$ gravity. At the beginning of Section-\ref{sec3}, we derived the well-known strong energy condition (SEC), and the null energy condition (NEC) in $f(R,Lm)$ gravity from Raychaudhuri equations for the test particle. Later on in Section-\ref{sec3},we obtained equivalent result by taking the transformation $p\rightarrow p^{eff}$ and $\rho \rightarrow \rho^{eff}$ into NEC and SEC of $f(R,L_m)$ that is $\rho + p \geq 0$ and $\rho +3p \geq0$, respectively . Also, we obtain the WEC and DEC in $f(R,L_m)$ gravity that is $\rho\geq0$ and $\rho- p\geq0$  from this transformation \cite{Brown/1993,Bertolami}. In Section-\ref{sec4} the energy conditions studied with the special model with $f(R,L_m)=e^{\alpha R}-\beta R L_m$ by using parameters of the jerk, the snap, and the deceleration parameters. Finally, gathering all the outcomes, we concluded in Section-\ref{sec5}.

\section{$f(R,L_m)$ gravity}\label{sec2}
Here, we have considered a well-motivated action for $f(R,L_m)$ gravity has been present in \cite{Harko/2010}. The action is given as
\begin{equation}\label{1}
S=\int\,f(R,L_m)\sqrt{-g}\,d^4x
\end{equation}

where $f(R,L_m)$ be a arbitrary function of $R$ and $L_m$. Here $R$ represent the Ricci scalar and $L_m$ represent the Lagrangian density of matter. The action of $f(R)$ with casual  matter-geometry coupling is redeem when $f(R,L_m)=\frac{1}{2}f_1(R)+G(L_m)f_2(R)$, where $f_i(R) (i=1,2)$ and $G(L_m)$ are general function of the Ricci scalar and the Lagrangian density of matter, respectively. 

 Varying the action \eqref{1} with respect to metric $g^{\alpha\beta}$, yield the field equation:
\begin{multline}\label{2}
f_R(R,L_m)R_{\alpha\beta}+(g_{\alpha\beta}\square-\bigtriangledown_\alpha\bigtriangledown_\beta)f_R(R,L_m)-\frac{1}{2}[f(R,L_m)\\
-f_{Lm}(R,L_m)L_m]g_{\alpha\beta}=\frac{1}{2}f_{Lm}(R,L_m)T_{\alpha\beta}
\end{multline}
where $\square=g^{\alpha\beta}\bigtriangledown_\alpha\bigtriangledown_\beta , f_R(R,L_m)=\frac{\partial f(R,L_m)}{\partial R}$ and $f_{L_m}(R,L_m)=\frac{\partial f(R,L_m)}{L_m}$. It has been suppose that $L_m$ only depends upon the metric tensor.

By definition, the matter energy-momentum tensor is given by
\begin{equation}\label{3}
T=-\frac{2}{\sqrt{-g}}\frac{\delta (\sqrt{-g} L_m)}{\delta g^{\alpha\beta}}
\end{equation}

The contraction of equation \eqref{2} gives 

\begin{multline}\label{4}
f_R(R,L_m)R+3\square f_R(R,L_m)-2[f(R,L_m)-f_{L_m}(R,L_m)L_m]
=\frac{1}{2}f_{L_m}(R,L_m)T
\end{multline}
where $T=T^\alpha _\beta$.

Now, applying covariant derivation to the equation \eqref{2}, one can write \eqref{2} as \cite{Harko/2010},
\begin{multline}\label{5}
\bigtriangledown^{\alpha}[f_R(R,L_m)R_{\alpha\beta}-\frac{1}{2}f(R,L_m)g_{\alpha\beta}
+(g_{\alpha\beta}\square-\bigtriangledown_\alpha\bigtriangledown_\beta)f_R(R,L_m)]\equiv 0.
\end{multline}
Furthermore, one can write the generalized conservation equation as
\begin{equation}\label{6}
\bigtriangledown^{\alpha}T_{\alpha\beta}=2\bigtriangledown^{\alpha}ln[f_{L_m}(R,L_m)]\frac{\partial L_m}{\partial g^{\alpha\beta}}.
\end{equation}

\section{Energy Conditions}\label{sec3}

\subsection{The Raychaudhuri equation}

Energy conditions plays vital role to understand the casual and geodesic behavior of spacetime. These conditions are derived from the well-known Raychaudhuri equation \cite{Rau/1955}. Energy conditions in $f(R,L_m)$ gravity can obtain by simply reviewing the Raychaudhuri equation.

The Raychaudhuri equation for congruence of time-like geodesics with the vector field $u^\alpha$ is defined as: 
\begin{equation}\label{7}
\frac{d\theta}{d\tau}=-\frac{1}{3}\theta^2 -\sigma_{\alpha\beta}\sigma^{\alpha\beta}+\omega_{\alpha\beta}\omega^{\alpha\beta}-R_{\alpha\beta}u^\alpha u^\beta,
\end{equation}
where $ \theta,\, R_{\alpha\beta},\,\omega_{\alpha\beta}$ and $\sigma_{\alpha\beta}$ represents the expansion parameter, the Ricci tensor, the rotation and  the shear associate with the congruence resp. While in the case of null geodesics, the Raychaudhuri equation is given by  

\begin{equation}\label{8}
\frac{d\theta}{d\tau}=-\frac{1}{2}\theta^2 -\sigma_{\alpha\beta}\sigma^{\alpha\beta}+\omega_{\alpha\beta}\omega^{\alpha\beta}-R_{\alpha\beta}k^\alpha k^\beta,
\end{equation}
here the null vector field is $k^\alpha$. From the above equations, it is clear that the Raychaudhuri equation is a purely geometric statement and independent of the gravity theory. In order to constrain the energy momentum tensor by the Raychaudhuri equation, the Ricci tensor helps to connect it with the motion equations of gravity. In case of GR, one can write equations \eqref{7} and \eqref{8} as (keeping in mind that gravity is always attractive)

\begin{equation}\label{9}
\text{SEC: } R_{\alpha\beta}u^\alpha u^\beta \geq 0                               
\end{equation}

\begin{equation}\label{10}
\text{NEC: }  R_{\alpha\beta}k^\alpha k^\beta \geq 0
\end{equation}
Using \eqref{9} and motion equations, one can obtain
\begin{equation}\label{11}
R_{\alpha\beta}u^\alpha u^\beta =\left(T_{\alpha\beta}-\frac{T}{2}g_{\alpha\beta} \right)u^\alpha u^\beta \geq 0
\end{equation}
where T is trace of energy-momentum tensor $T_{\alpha\beta}$.  Now, we  consider the fluid description of spacetime as a perfect fluid with energy density $\rho$ and pressure $p$, 
\begin{equation}\label{12}
T_{\alpha\beta}=(\rho+p)u_\alpha u_\beta -pg_{\alpha\beta}.
\end{equation}
Now, one can recover SEC from \eqref{11} in GR theory
\begin{equation}\label{13}
\rho +3p \geq 0.
\end{equation}
Also, from equation \eqref{10} and Einstein's field equations, we get  
\begin{equation}\label{14}
T_{\alpha\beta}k^\alpha k^\beta \geq 0.
\end{equation}
Now, from \eqref{12}, the NEC reads as 
\begin{equation}\label{15}
\rho+p\geq 0.
\end{equation}

\subsection{ Energy Condition in $ f(R,L_m) $ gravity} 

In analogy with GR, we can write the field equation \eqref{2} as: 

\begin{equation}\label{16}
G_{\alpha\beta}\equiv R_{\alpha\beta}-\frac{1}{2}g_{\alpha\beta}R=T^{eff}_{\alpha\beta},
\end{equation}
where $T^{eff}_{\alpha\beta}$ is the effective energy-momentum tensor and it can be represented as
\begin{multline}\label{17}
T^{eff}_{\alpha\beta}=\frac{1}{f_R(R,L_m)}\left\lbrace\frac{1}{2}g_{\alpha\beta}[f(R,L_m)-Rf_R(R,L_m)]-(g_{\alpha\beta}\square -\bigtriangledown_\alpha \bigtriangledown_\beta)f_R(R,L_m)
\right\rbrace\\
+\frac{1}{f_R(R,L_m)}\left\lbrace\frac{1}{2}f_{L_m}(R,L_m)T_{\alpha\beta}\right\rbrace
\end{multline}

After contracting equation \eqref{17}, we get 
\begin{multline}\label{18}
T^{eff}=\frac{1}{f_R(R,L_m)}\left\lbrace 2[f(R,L_m)-Rf_R(R,L_m)]-3\square 
f_R(R,L_m)+2f_{L_m}(R,L_m)L_m\right\rbrace\\+\frac{1}{f_R(R,L_m)}\left\lbrace \frac{1}{2}f_{L_m}(R,L_m)T\right\rbrace
\end{multline}

where $T=g^{\alpha\beta} T_{\alpha\beta}$. Hence corresponding Ricci tensor reads
\begin{equation}\label{19}
R_{\alpha\beta}=T^{eff}_{\alpha\beta}-\frac{1}{2}g_{\alpha\beta}T^{eff}.
\end{equation}
For attractive gravity, we need an additional condition which is different from Raychaudhuri equation as 
\begin{equation}\label{20}
\frac{f_{L_m}(R,L_m)}{f_R(R,L_m)}> 0
\end{equation}

The flat and isotropic metric is chosen as: 
\begin{equation}\label{21}
ds^2=dt^2-a^2(t)dx^{2}_{3}
\end{equation}
where $dx^2 _3$ carry the spatial part of the metric and $a(t)$ represents the scale factor. We can obtain $R=-6(2H^2+\dot{H})$, here $H=\frac{\dot{a(t)}}{a(t)}$ is the Hubble expansion parameter, and $\Gamma^0 _{\alpha\beta}=a(t)\dot{a(t)}\delta_{\alpha\beta} (\alpha\beta \neq 0)$, which are components of the affine connection.
  
Now, from condition \eqref{9} and \eqref{19}, the SEC  for $f(R,L_m)$ gravity reads 

\begin{equation}\label{22}
T^{eff}_{\alpha\beta} u^\alpha u^\beta-\frac{1}{2}T^{eff}\geq 0.
\end{equation}
Then equation \eqref{22}  with \eqref{12} and \eqref{20} turns into (where we have used the condition $g_{\alpha\beta} u^\alpha u^\beta=1$ ) 
\begin{multline}\label{23}
\rho+3p-\frac{2}{f_{L_m}(R,L_m)}[f(R,L_m)-R f_R(R,L_m)]\\
+\frac{6}{f_{L_m}(R,L_m)}[f_{RRR}(R,L_m)\dot{R}^2+f_{RR}(R,L_m)\ddot{R}
+Hf_{RR}(R,L_m)\dot{R}-2L_m \geq 0,
\end{multline} 
here,$^{`.'}=\frac{d}{dt}$.
The NEC for $f(R,L_m)$ gravity reads:

\begin{equation}\label{24}
T^{eff}_{\alpha\beta}k^\alpha k^\beta \geq 0
\end{equation}
by the fallowing same path as the SEC, the above relationship turns into as fallow: 
\begin{equation}\label{25}
\rho+p+\frac{2}{f_{L_m}(R,L_m)}[F_{RRR}(R,L_m)\dot{R}^2+f_{RR}(R,L_m)\ddot{R}]\geq 0.
\end{equation}

Now, following Raychaudhuri equation and by taking transformations $\rho\longrightarrow \rho^{eff}$ and $p\longrightarrow p^{eff}$, we can directly derived the expressions for SEC and NEC. As a result, we find equivalent results $\rho+p\geq 0$ and $\rho+3p\geq 0$ for NEC and SEC,  respectively. If one expand this approach to $\rho\geq 0$ and $\rho-p\geq 0$, then one can find WEC and DEC in $f(R,L_m)$ gravity.

By using the equation \eqref{17} and \eqref{21}, the effective energy density and the effective pressure can obtained as 

\begin{multline}\label{26}
\rho^{eff}=\frac{1}{f_R(R,L_m)}\lbrace\frac{1}{2}[f(R,L_m)-Rf_R(R,L_m)]-3Hf_{RR}(R,L_m)\dot{R}+\frac{1}{2}f_{L_m}(R,L_m)L_m\\
+\frac{1}{2}f_{L_m}(R,L_m)\rho\rbrace,
\end{multline}
\begin{multline}\label{27}
p^{eff}=\frac{1}{f_R(R,L_m)}\left\lbrace\frac{1}{2}[Rf_R(R,L_m)-f(R,L_m)]+f_{RRR}(R,L_m)\dot{R}^2+f_{RR}(R,L_m)\ddot{R}\right\rbrace
 \\
+\frac{1}{f_R(R,L_m)}\left\lbrace 3Hf_{RR}(R,L_m)\dot{R}-\frac{1}{2}f_{L_m}(R,L_m)L_m+\frac{1}{2}f_{L_m}(R,L_m)p \right\rbrace
\end{multline}
Then, the DEC in $f(R,L_m)$ gravity can be written as
\begin{multline}\label{28}
\rho-p+\frac{2}{f_{L_m}(R,L_m)}[f(R,L_m)-R f_R(R,L_m)]\\
-\frac{2}{f_{L_m}(R,L_m)}[f_{RRR}(R,L_m)\dot{R}^2+f_{RR}(R,L_m)\ddot{R}+3Hf_{RR}(R,L_m)\dot{R}]+2L_m \geq 0,
\end{multline}

and, the WEC in $f(R,L_m)$ gravity take the following form
\begin{equation}\label{29}
\rho+\frac{1}{f_{L_m}(R,L_m)}[f(R,L_m)-R f_R(R,L_m)]-\frac{6}{f_{L_m}(R,L_m)}Hf_{RR}(R,L_m)\dot{R}+L_m \geq 0.
\end{equation}
Now, in the further study one can use the above energy conditions to explore some cosmological applications in $f(R, L_m)$ gravity.

\section{constraint on $f(R,L_m)$ gravity}\label{sec4}

In this section, our aim is to constraint the model parameters in order to discuss the present cosmological scenario. For this purpose, we are going to use the cosmographic parameters such as Hubble parameter $H$, deceleration parameter ($q$), jerk parameter ($j$) and snap parameter ($s$). Furthermore, One can write the Ricci scalar $R$ and its derivatives in terms of cosmographic parameters as
\begin{align}\label{30}
R=-6H^2 (1-q)  
\end{align}
\begin{equation}\label{31}
 \dot{R}=-6H^3(j-q-2)
\end{equation}
\begin{equation}\label{32}
\ddot{R}=-6H^4(s+8q+q^2+6)
\end{equation}

where 
\begin{equation}\label{33}
q=\frac{1}{H^2}\frac{\ddot{a}}{a}, j=\frac{1}{H^3}\frac{\ddot{a}}{a}     \text{ and }s=\frac{1}{H^4}\frac{\ddot{a}}{a}.
\end{equation}

To proceed further and to explore the cosmological scenarios, we presume the Lagrangian $f(R,L_m)$ as
\begin{equation}\label{34}
f(R,L_m)= e^{\alpha R}-\beta R L_m
\end{equation}
 where $\alpha,\, \beta $ are non-zero constants.

Then, the energy condition \eqref{23}, \eqref{25}, \eqref{28}, and \eqref{29} can be rewritten as \\


\begin{multline}\label{35}
\text{\textbf{SEC:} }\rho+3p+\frac{2e^{6\alpha H^2(q-1)}}{\beta (q-1)}\times\\
\left[\frac{1}{6H^2}-\alpha (q-1)-3\alpha^2 H^2[6\alpha H^2(j-q-2)^2-(s+j+7q+q^2+4)]\right]-2L_m\geq 0
\end{multline}

\begin{equation}\label{36}
\text{\textbf{NEC:} }\rho+p-\frac{2\alpha^2H^2e^{6\alpha H^2(q-1)}}{\beta (q-1)}\left[6\alpha H^2(j-q-2)^2-(s+8q+q^2+6)\right]\geq 0
\end{equation}

\begin{multline}\label{37}
\text{\textbf{DEC:} }\rho-p-\frac{2e^{6\alpha H^2(q-1)}}{\beta (q-1)}\times \\
\left[\frac{1}{6H^2}-\alpha (q-1)-\alpha^2 H^2[6\alpha H^2(j-q-2)^2-(s+3j+5q+q^2)]\right]+2L_m\geq 0
\end{multline}

\begin{equation}\label{38}
\text{\textbf{WEC:} }\rho-\frac{e^{6\alpha H^2(q-1)}}{\beta (q-1)}\left[\frac{1}{6H^2}-\alpha(q-1)+6\alpha^2 H^2(j-q-2)\right]+L_m\geq 0
\end{equation}

To simplify our discussion, we rewrite the ECs with the present time as

\begin{align}
\label{39}
\text{\textbf{SEC:} }\rho+3p-2L_m\geq A(\alpha,\beta),
\end{align}
\begin{multline*}\text{here, }A(\alpha,\beta)=-\frac{2e^{6\alpha H_0^2(q_0-1)}}{\beta (q_0-1)}\times\\
\left[\frac{1}{6H_0^2}-\alpha (q_0-1)-3\alpha^2 H_0^2[6\alpha H_0^2(j_0-q_0-2)^2-(s_0+j_0+7q_0+q^2+4)]\right]
\end{multline*}

\begin{equation}\label{40}
\text{\textbf{NEC:} }\rho+p\geq B(\alpha,\beta)
\end{equation}
where, $B(\alpha,\beta)=\frac{2\alpha^2H_0^2e^{6\alpha H_0^2(q_0-1)}}{\beta (q_0-1)}\left[6\alpha H_0^2(j_0-q_0-2)^2-(s_0+8q_0+q_0^2+6)\right]$

\begin{equation}\label{41}
\text{\textbf{DEC: }}\rho-p+2L_m\geq C(\alpha,\beta)
\end{equation}
where \begin{multline}C(\alpha,\beta)=\frac{2e^{6\alpha H_0^2(q_0-1)}}{\beta (q_0-1)}\times \\
\left[\frac{1}{6H_0^2}-\alpha (q_0-1)-\alpha^2 H_0^2[6\alpha H_0^2(j_0-q_0-2)^2-(s_0+3j_0+5q_0+q_0^2)]\right]\end{multline}
 
\begin{equation}\label{42}
\text{\textbf{WEC:}  }\rho+L_m\geq D(\alpha,\beta)
\end{equation}
where, $D(\alpha,\beta)=\frac{e^{6\alpha H_0^2(q_0-1)}}{\beta (q_0-1)}\left[\frac{1}{6H_0^2}-\alpha(q_0-1)+6\alpha^2 H_0^2(j_0-q_0-2)\right]$

To discuss the cosmological model's viability, we take the present values of cosmological parameters such as $H_0=67.9,\, q_0=-0.664 ,\, j_0=1.223 ,\, s_0= 0.394$ \cite{Planck,Capo/2019}. As we discussed previous, currently our universe is showing accelerated expansion and for that SEC has to violate of a cosmological model \cite{Planck}.

\subsection{Case-1: $L_m=-\rho$}

In this subsection, our aim is to constraint the model parameters $\alpha,\,\, \beta$. For this purpose, we consider the observational constraint values of the cosmographic parameters such as $H_0,\, q_0,\, j_0,\, s_0$. Moreover, $\rho^{eff}$ should be positive throughout the evolution of the universe. Keeping this in mind, from equation \eqref{42}, we find that $-D(\alpha ,\beta)\geq 0$ for $\alpha<0,\, \beta<0$. With this range of parameters, we also observe that $A(\alpha,\beta)<0,\, B(\alpha,\beta)<0 $ and $C(\alpha,\beta)>0$. These inequalities suggest that SEC and WEC violates whereas DEC satisfies. The behavior of our model align with the models presented in \cite{Whin} and shows the accelerated expansion of the universe in the present time.

\subsection{Case-II: $L_m=p$}

Here, in this subsection, we choose the matter Lagrangian $L_m=p$. With the values of cosmographic parameters and following the procedure discussed in first case, we are able to constraint the model parameters $\alpha,\, \beta$. For $\alpha> 0$ and $\beta> 0$, we find that $A(\alpha,\beta)<0,\, B(\alpha,\beta)<0,\, C(\alpha,\beta)>0$ and $D(\alpha,\beta)>0$. The results for this case is also an agreement with the current accelerated expansion of the universe.

\section{Conclusion}\label{sec5}

In the last few decades, observational cosmology has been growing rapidly and motivating researchers to go beyond the standard description of gravity, i.e., Einstein's GR. In this regard, we presume a modified Hilbert-Einstein action, where the Lagrangian $f(R)$ is replaced by $f(R, L_m)$. However, one of the crucial roles is to test their self-consistencies in energy conditions. The main motivation for working in this theory is to discuss the current status of the universe through the cosmological model and to check its' compatibility with the casual and geodesic structure of space. And, it is well-known that energy conditions play an important role in checking such type test. In this study, we have discussed the energy conditions following the formulation of the energy conditions in the framework of modified theories of gravity \cite{capo/2014, capo/2015, capo/2018} for $f(R,L_m)$ gravity. This procedure allows extra degrees of freedom, which emerges concerning general relativity (GR). Further, we grouped them as an effective energy-momentum tensor($T^{eff}$). In proceeding, we derived the strong, the weak, the dominant, and the null energy conditions in the context of $f(R,L_m)$ gravity in the GR analogy. Although we have followed the GR analogy, the energy conditions meaning can be totally different from GR, and the geodesic structure, casual structure, and gravitational interaction may be altered. This suggests that the physical meaning of the modified theories of gravity depends on the energy conditions \cite{capo/2009}. We also derived all the energy conditions for a specific $f(R, L_m)$ Lagrangian has two parameters $\alpha$, and $\beta$. Moreover, we discussed energy conditions with two different choices of matter Lagrangian $L_m$ such as for $L_m=-\rho$ and $L_m=p$. In GR, without considering a cosmological constant, a non-positive contribution in the Raychaudhuri equation is generally interpreted as the manifestation of the attractive character of gravity. In particular, strong energy condition (SEC) has to be satisfied to maintain the attractive character of gravity. But, in the case of the modified theory of gravity, one of the energy conditions may positively contribute to the Raychaudhuri equation. This fact opens the possibility to explain the accelerated expansion of the universe or the repulsive behaviour of modified theories of gravity \cite{santos/2017}. SEC needs to be violated to achieve the repulsive behaviour in the context of modified theories of gravity \cite{santos/2017}. Keeping this in mind and using the recent observational values of cosmographic parameters, we can constrain the model parameters $\alpha$ and $\beta$. In case $L_m=-\rho$, we observed that for $\alpha<0,\, \beta<0$, our model shows the accelerated expansion of the universe. But, in case of $L_m=p$, model shows accelerated behaviour for $\alpha>0,\, \beta>0$. 

These previous findings enabled us to test the viability of various families of $f(R,L_m)$ gravity models, paving the way for a comprehensive description of gravity compatible in the dark energy era. Another intriguing fact is that our free parameters have a lot of leeways, allowing for several testable scenarios for $f(R,L_m)$ gravity. Furthermore, exploring the coupling of $f(R, L_m)$ with the scalar fields would be interesting, looking into cosmological parameters constraints or possible analytical models.

\section*{Acknowledgements}

L.V.J. acknowledges University Grant Commission (UGC), Govt. of India, New Delhi, for awarding JRF (NTA Ref. No.: 191620024300). S.M. acknowledges Department of Science \& Technology (DST), Govt. of India, New Delhi, for awarding INSPIRE Fellowship (File No. DST/INSPIRE Fellowship/2018/IF180676). We are very much grateful to the honorable referee and to the editor for the
illuminating suggestions that have significantly improved our work in terms
of research quality, and presentation.


\begin{thebibliography}{90}

\bibitem{Riess/1998}A. G. Riess  et al., Observational Evidence from Supernovae for an Accelerating Universe and a Cosmological Constant, \textit{ApJ} \,\
\textbf{116}  (1998), 1009.

\bibitem{Perlmutter/1999}S. Perlmutter  et al., Measurements of $\Omega$ and $\Lambda$ from 42 High-Redshift Supernovae, \textit{Astrophys.J.}\,\
\textbf{517} (1999), 565.

\bibitem{Teqmark/2004}M. Tegmark  et al., Cosmological parameters from SDSS and WMA, \textit{ Phys. Rev. D.}\,\
\textbf{69} (2004), 103501.

\bibitem{Teqmark/2004a}M. Tegmark  et al., The Three-Dimensional Power Spectrum of Galaxies from the Sloan Digital Sky Survey,  \textit{Astrophys. J.}\,\
\textbf{606} (2004), 702.

\bibitem{Wetterich/1988}C. Wetterich, Cosmology and the fate of dilatation symmetry, \textit{ Nucl. Phys. B.}\,\
\textbf{302} (1988), 668.

\bibitem{Amendola/2000}L. Amendola, Coupled quintessence, \textit{Phys. Rev. D.}\,\
\textbf{62} (2000) 043511.

\bibitem{calwell/2002}R. R. Caldwell, A phantom menace? Cosmological consequences of a dark energy component with super-negative equation of state, \textit{ Phys. Lett. B.}\,\
\textbf{545} (2002), 23 .

\bibitem{Brans}C. Brans  and R. H. Dicke, Mach's Principle and a Relativistic Theory of Gravitation, \textit{Phys. Rev.}\,\
\textbf{124} (1961), 925 .

\bibitem{Fara}V. Faraoni, \textit{Cosmology in Scalar-Tensor Gravity} (Springer, Dordrecht, 2004).

\bibitem{Dvali}G. Dvali et al., 4D gravity on a brane in 5D Minkowski space, \textit{ Phys. Lett. B} \,\
\textbf{485} (2000), 208 .

\bibitem{Jacobson}T. Jacobson  and D. Mattingly, Gravity with a dynamical preferred frame, \textit{ Phys. Rev. D}\,\
\textbf{64} (2001), 024028 .

\bibitem{Maattens}R. Maartens, Brane-World Gravity, \textit{Living Rev. Rel}.\,\
\textbf{7} (2004), 7.

\bibitem{Beken}J. D. Bekenstein, Relativistic gravitation theory for the modified Newtonian dynamics paradigm, \textit{Phys. Rev. D}\,\
\textbf{70} (2004), 083509.

\bibitem{Njiroi/2007}S. Nojiri  and S. D. Odintsov, Introduction to modified gravity and gravitational alternative fro dark energy, \textit{Int. J. Geom. Methods Mod. Phys.}\,\
\textbf{4} (2007), 115.

\bibitem{Sotiriou/2010}T. P. Sotiriou and V. Faraoni, $f(R)$ theories of gravity \textit{Rev. Mod. Phys.}\,\
\textbf{82} (2010), 451 .

\bibitem{Capo/2002}S. Capozziello, Curvature quintessence, \textit{Int. J. Mod. Phys. D}\,\
\textbf{11} (2002), 483.

\bibitem{Nojiri/2008}S. Nojiri and S. D. Odintsov, Modified Gravity with ln R Terms and Cosmic Acceleration, \textit{Gen. Rel. Grav.}\,\
\textbf{36} (2004), 1765.

\bibitem{Nojiri/2003}S. Nojiri  and S. D. Odintsov, Modified gravity with negative and positive powers of curvature: Unification of inflation and cosmic acceleration, \textit{Phys. Rev. D}\,\
\textbf{68} (2003), 123512.

\bibitem{Nojiri/2005}S. Nojiri and S. D. Odintsov, Modified Gauss–Bonnet theory as gravitational alternative for dark energy, \textit{Phys. Lett. B}\,\
\textbf{631} (2005), 1 .

\bibitem{Berto/2007} O. Bertolami, C. G. Boehmer, T. Harko  and F. S. N. Lobo, Extra force in $f(R)$ modified theories of gravity, \textit{Phys. Rev. D}\,\
\textbf{75} (2007), 104016.

\bibitem{Harko/2008}T. Harko, Modified gravity with arbitrary coupling between matter and geometry, \textit{ Phys. Lett. B}\,\
\textbf{669} (2008), 376.

\bibitem{Harko/2010}T. Harko and F. S. N. Lobo, $f(R,L_m)$ gravity, \textit{ Eur. Phys. J. C}\,\
\textbf{70} (2010), 373.

\bibitem{Kung}J. H. Kung, Strong energy condition in $R+R^2$ gravity  \textit{Phys. Rev. D}\,\
\textbf{53} 1996), 3017.

\bibitem{Carroll}S. Carroll, \textit{Spacetime and Geometry: An Introduction to General Relativity} (Addison-Wesley, New York, 2004).

\bibitem{Allemandi}G. Allemandi et al., Dark energy dominance and cosmic acceleration in first-order formalism, \textit{ Phys. Rev. D}\,\
\textbf{72} 2005, 063505.

\bibitem{Berg/2006}S. E. P. Bergliaffa, Constraining $f(R)$ theories with the energy conditions,  \textit{Phys. Lett. B }\,\
\textbf{642} (2006), 311.

\bibitem{Santos/2007}J. Santos, J. S. Alcaniz, M. J. Reboucas and F. C. Carvalho, Energy conditions in $f(R)$ gravity \textit{Phys. Rev. D}\,\
\textbf{76} (2007), 083513.

\bibitem{Berto/2009}O. Bertolami  and M. C. Sequeira, Energy conditions and stability in $f(R)$ theories of gravity with nonminimal coupling to matter, \textit{ Phys. Rev. D}\,\
\textbf{79} (2009), 104010.

\bibitem{Wang/2010}J. Wang, W. Ya-Bo, Y. X. Guo, W. Q. Yang and L. Wang, Energy conditions and stability in generalized $f(R)$ gravity with arbitrary coupling between matter and geometry \textit{ Phys. Lett. B} \,\
\textbf{689} (2010), 133.

\bibitem{Gracia/2011}M. Garcia, T. Harko, F. S. N. Lobo and J. P. Mimoso, Energy conditions in modified Gauss-Bonnet gravity, \textit{Phys. Rev. D }\,\
\textbf{83} (2011), 104032.

\bibitem{wald}R. M. Wald, \textit{General Relativity} (University of Chicago Press, Chicago, 1984).

\bibitem{Rau/1955}A. Raychaudhuri, Relativistic Cosmology. I,  \textit{Phys. Rev. D}\,\
\textbf{98} (1955), 1123.


\bibitem{Moraes/2019}P. H. R. S.  Moraes, et al., A Cosmological Scenario from the Starobinsky Model within the $f(R,T)$ Formalism \textit{Adv. Astron.},  \textbf{2019} (2019), 8574798.

\bibitem{Bamba/2017}K. Bamba  et al., Energy conditions in modified f(G) gravity, \textit{Gen. Relativ. Grav.},  \textbf{49} (2017), 112.

\bibitem{Sharif/2013}M. Sharif, M. Zubair, Energy conditions in $f(R,T, R_{\mu\nu} T_{\mu\nu})$ gravity, \textit{J. High Energy Physics}, \textbf{2013} (2013), 79.

\bibitem{Liu/2012}D. Liu, M. J. Rebouc, Energy conditions bounds on $f(T)$ gravity, \textit{Phys. Rev D,} \textbf{86} (2012), 083515.

\bibitem{Atazadeh/2014}K. Atazadeh, F. Darabi, Energy conditions in $f(R,G)$ gravity,  \textit{Gen. Relativ. Grav.},  \textbf{46} (2014),  1664.

\bibitem{Sharif/2016}M. Sharif, A. Ikram, Energy conditions in $f(G,T)$ gravity, \textit{ Eup. Phys. J. C}, \textbf{76} (2016), 640.

\bibitem{Yousaf/2018}Z. Yousaf et al., Energy conditions in higher derivative $f(R,\square R,T)$ gravity, \textit{Int. J. Geom. Methods Mod. Phys.},  \textbf{15} (2018), 1850146.

\bibitem{rl1}J. Wang and K. Liao, Energy conditions in $f(R, L_m)$ gravity \textit{Class. Quantum Grav.} \textbf{29} (2012), 215016.

\bibitem{rl2}Y. Wu  et al., Constraints of energy conditions and DK instability criterion on $f(R, L_m)$ gravity models, Mod. Phys. Lett. A\textbf{29} (2014), 1450089.


\bibitem{Mandal}S. Mandal, P. K. Sahoo, J. R. L. Santos, Energy conditions in $f(Q)$ gravity \textit{Phys. Rev. D}, \, \textbf{102} (2020), 024057.

\bibitem{Simran}S. Arora, J. R. L. Santos, P. K. Sahoo, Constraining $f(Q,T)$ gravity from energy conditions
, \textit{ Phys. Dark. Univ.} \textbf{31} (2021), 100790.

\bibitem{Sharif/2013}M.
Sharif, M. Zubair, Thermodynamics in Modified Gravity with Curvature Matter Coupling, 
\textit{Adv. High Energy Phys.} \textbf{2013} (2013), 947898.

\bibitem{Brown/1993}J. D. Brown, Action functionals for relativistic perfect fluids, \textit{Class. Quantum Grav.}\,\
\textbf{10} (1993), 1579.

\bibitem{Bertolami}O. Bertolami, T. Harko, F. S. N. Lobo and J. Paramos, Non-minimal curvature-matter couplings in modified gravity, \textit{arXiv:0811.2876.}

\bibitem{Planck} N. Aghanim et al., Planck 2018 results. VI. Cosmological parameters, \textit{A \& A} \textbf{641} (2020) 67. 

\bibitem{Capo/2019}S. Capozziello et al., Extended gravity cosmography, \textit{Int. J. Mod. Phys. D}, \textbf{28} (2019),  1930016.

\bibitem{Whin}A. W. Whinnett, D.F. Torres, A New Strong-Field Effect in Scalar Tensor Gravity: Spontaneous Violation of the Energy Conditions \textit{ The Astrophysical Journal,} \textbf{603} (2004), L133 ; G. Calcagni, \textit{Classical and Quantum Cosmology} (Springer Switzerland, 2017).

\bibitem{capo/2014}S. Capozziello, F. S. N. Lobo and J. P. Mimoso, Energy conditions in modified gravity,\textit{Phys. Lett. B} \textbf{730} (2014), 280.

\bibitem{capo/2015}S. Capozziello, F. S. N. Lobo and J. P. Mimoso, Generalized energy conditions in extended theories of gravity\textit{Phys. Rev. D} \textbf{91} (2015) 124019.

\bibitem{capo/2018}S. Capozziello, S. Nojiri and S. D. Odintsov,The role of energy conditions in f(R) cosmology,\textit{Phys. Lett. B} \textbf{781} (2018) 99.

\bibitem{capo/2009} S. Capozziello, S. Vignolo, on the well-formulation of the initial value problem of metric-affine f(R) Gravity ,\textit{Int. J. Geom. Methods Mod. Phys} \textbf{6} (2009) 985.

\bibitem{santos/2017}C.S. Santos et al., Strong energy condition and the repulsive character of f(R) gravity, \textit{Gen. Realtiv. Gravit.} \textbf{49} (2017), 50.




\end{thebibliography}
\end{document}